\documentclass{article}
\usepackage{times}
\usepackage{amsfonts}
\usepackage{graphicx}
\usepackage[pdfmark]{hyperref}
\begin{document}
\noindent
{\Large EMERGENCE FROM IRREVERSIBILITY}
\vskip1cm
\noindent
{\bf  P. Fern\'andez de C\'ordoba$^{1,a}$, J.M. Isidro$^{1,b}$ and Milton H. Perea$^{1,2,c}$}\\
${}^{1}$Instituto Universitario de Matem\'atica Pura y Aplicada,\\ Universidad Polit\'ecnica de Valencia, Valencia 46022, Spain\\
${}^{2}$Departamento de Matem\'aticas y F\'{\i}sica, Universidad Tecnol\'ogica\\ del Choc\'o, Colombia\\
${}^{a}${\tt pfernandez@mat.upv.es}, ${}^{b}${\tt joissan@mat.upv.es}\\
${}^{c}${\tt milpecr@posgrado.upv.es}  \\
\vskip.5cm
\noindent
{\bf Abstract} The emergent nature of quantum mechanics is shown to follow from a precise correspondence with the classical theory of irreversible thermodynamics.

\section{Introduction}\label{scuno}

The aim of this talk\footnote{Work partially based on ref. \cite{XNOI3} by some of the present authors (P. F. de C. and J.M.I.), presented by J.M.I. at the {\it Sixth International Workshop DICE 2012: Spacetime--Matter--Quantum Mechanics:
from the Planck scale to emergent phenomena (Castiglioncello, Italy, September 2012).}} is to establish a correspondence between quantum mechanics, on the one hand, and the classical thermodynamics of irreversible processes, on the other. This we do in order to provide an independent proof of the statement that {\it quantum mechanics is an emergent phenomenon}\/. The emergent aspects of quantum mechanics have been the subject of a vast literature; a very incomplete list of refs. would include \cite{XNOI1, XNOI2, XNOI3, BLASONE0, BLASONE00,  BLASONE1, BLASONE2, BLASONE3, XCARROLL, XELZE1, XELZE2, XELZE3, XELZE4, XGROESSING1, XGROESSING2, XTHOOFT1, XTHOOFT2, XTHOOFT3, XTHOOFT4, XTHOOFT5, KHRENNIKOV1, KHRENNIKOV2, XSTABILE, XSMOLIN, WETTERICH}.

\section{Basics in irreversible thermodynamics}\label{scdos}

We first summarise, for later use, some basic elements of the classical thermodynamics of irreversible processes in the linear regime \cite{XONSAGER}.

Let a thermodynamical system be given, deviating only slightly from equilibrium. Assume that its entropy $S$ depends on $N$ extensive variables $y^1, \ldots, y^N$, so we can write $S=S(y^1,\ldots, y^N)$. The tendency of the system to seek equilibrium is measured by the {\it thermodynamic forces}\/ $Y_k$, defined to be the components of the gradient of the entropy:
\begin{equation}
Y_k:=\frac{\partial S}{\partial y^k}.
\label{fuerzas}
\end{equation}
Now our system is away from equilibrium, but not too far away, so we can assume linearity between the fluxes $\dot y^k$ and the forces $Y_j$:
\begin{equation}
\dot y^i:=\frac{{\rm d}y^i}{{\rm d}\tau}=\sum_{j=1}^NL^{ij}Y_j,\qquad  Y_i=\sum_{j=1}^NR_{ij}\dot y^j, \qquad R_{ij}=(L^{ij})^{-1}.
\label{flujos}
\end{equation}
We use $\tau$ to denote thermodynamical time, and we suppose the above relation between forces and fluxes to be invertible. A well--known result is Onsager's reciprocity theorem: the matrix $L$ is symmetric,
\begin{equation}
L^{ij}=L^{ji}.
\label{reciprocidad}
\end{equation}
By (\ref{flujos}), the rate of entropy production can be written either as a quadratic form in the fluxes, or as a quadratic form in the forces:
\begin{equation}
\dot S=\sum_{j=1}^N\frac{\partial S}{\partial y^j}\dot y^j=\sum_{j=1}^NY_j\dot y^j=\sum_{i,j=1}^NR_{ij}\dot y^i\dot y^j=\sum_{i,j=1}^NL^{ij}Y_iY_j.
\label{tasa}
\end{equation}
We can Taylor expand the entropy $S$ around equilibrium and truncate the series at second order, to find
\begin{equation}
S=S_0-\frac{1}{2}\sum_{i,j=1}^Ns_{ij}y^iy^j+\ldots,
\label{teilor}
\end{equation}
where the matrix $s_{ij}=-\partial^2 S/\partial y^i\partial y^j\vert_0$ (the negative Hessian evaluated at equilibrium) is positive definite. This truncation has the consequence that fluctuations around equilibrium are Gaussian. Indeed, by Boltzmann's principle, the probability $P(y^1,\ldots, y^N)$ of finding the values $y^1,\ldots, y^N$ of the extensive variables is given by
\begin{equation}
P(y^1,\ldots, y^N)=Z^{-1}\exp\left(\frac{S}{k_B}\right)=Z^{-1}\exp\left(-\frac{1}{2k_B}\sum_{i,j=1}^Ns_{ij}y^iy^j\right),
\label{volman}
\end{equation}
where $Z$ is a normalisation factor.

{}For simplicity we set $N=1$ in all that follows. Our aim is to calculate the probability of any path $y=y(\tau)$ in the thermodynamical configuration space.
A {\it cumulative}\/ distribution function $F_n\left({y_1\atop \tau_1}{\ldots\atop\ldots}{y_n\atop \tau_n}\right)$ is defined such that it yields the probability that the thermodynamical path $y(\tau)$ lie below the barriers $y_1, \ldots, y_n$ at times $\tau_1<\tau_2<\ldots< \tau_n$:
\begin{equation}
F_n\left({y_1\atop \tau_1}{\ldots\atop\ldots}{y_n\atop \tau_n}\right):=P\left(y(\tau_k)\leq y_k,\; k=1,\ldots, n\right).
\label{cumulaty}
\end{equation}
A stationary process is defined to be one such that $F_n$ is invariant under time shifts $\delta\tau$:
\begin{equation}
F_n\left({y_1\atop \tau_1}{\ldots\atop\ldots}{y_n\atop \tau_n}\right)=
F_n\left({y_1\atop \tau_1+\delta\tau}{\ldots\atop\ldots}{y_n\atop \tau_n+\delta\tau}\right).
\label{estacionar}
\end{equation}
In other words, the system that has been left alone long enough that any initial conditions have been forgotten.  An {\it unconditional}\/ probability density function $f_n\left({y_1\ldots y_n}\atop{\tau_1\ldots \tau_n}\right)$ is defined, such that the product
\begin{equation}
f_n\left({y_1\ldots y_n}\atop{\tau_1\ldots \tau_n}\right){\rm d}y_1\cdots{\rm d}y_n
\label{densita}
\end{equation}
measures the probability that a thermodynamical path $y=y(\tau)$ pass through a gate of width ${\rm d}y_k$ at instant $\tau_k$, for all $k=1, \ldots n$.
Similarly, the {\it conditional}\/ probability density function $f_1\left({y_k\atop \tau_k}{\Big \vert}{y_{k-1}\atop \tau_{k-1}}\right)$ is such that the product
\begin{equation}
f_1\left({y_k\atop \tau_k}{\Big \vert}{y_{k-1}\atop \tau_{k-1}}\right){\rm d}y_k\,{\rm d}y_{k-1}
\label{condizionale}
\end{equation}
gives the probability that $y=y(\tau)$ pass through ${\rm d}y_k$ at $\tau_k$, given that it passed through ${\rm d}y_{k-1}$ at $\tau_{k-1}$.
Finally a Markov process is defined to be {\it one that has a short memory}\/ or, more precisely, one such that its cumulative, conditional probability function satisfies
\begin{equation}
F_1\left({y_{n+1}\atop \tau_{n+1}}{\Big \vert}{y_{1}\atop \tau_1}{\ldots\atop\ldots}{y_{n}\atop \tau_n}\right)=
F_1\left({y_{n+1}\atop \tau_{n+1}}{\Big\vert}{y_{n}\atop \tau_n}\right).
\label{marcof}
\end{equation}

One can prove that, for a Markov process, the following factorisation theorem holds \cite{XONSAGER}:
\begin{equation}
f_n\left({y_{1}\ldots y_{n}}\atop{\tau_1\ldots \tau_n}\right)=f_1\left({y_{n}\atop \tau_n}{\Big\vert}{y_{n-1}\atop \tau_{n-1}}\right)
\cdots f_1\left({y_{2}\atop \tau_{2}}{\Big \vert}{y_{1}\atop \tau_{1}}\right)\, f_1\left({y_{1}\atop \tau_1}\right).
\label{marcuf}
\end{equation}
Interesting about this factorisation theorem is the fact that $f_1\left({y_{1}\atop \tau_1}\right)$ is known from Boltzmann's principle. Therefore, by stationarity, all we need to know is
\begin{equation}
f_1\left({y_{2}\atop \tau + \delta\tau}{\Big\vert}{y_{1}\atop \tau}\right),
\label{puerta}
\end{equation}
and solving the $n$--gate problem $f_n\left({y_{1}\ldots y_{n}}\atop{\tau_1\ldots \tau_n}\right)$ nicely reduces to solving the 2--gate problem $f_1\left({y_{2}\atop \tau + \delta\tau}{\Big\vert}{y_{1}\atop \tau}\right)$.

Now, under the assumption that our irreversible thermodynamical processes is stationary, Markov and Gaussian, the conditional probability density (\ref{puerta}) has been computed in \cite{XONSAGER}, with the result
\begin{equation}
f_1\left({y_{2}\atop \tau+\delta\tau}{\Big\vert}{y_{1}\atop \tau}\right)
=\frac{1}{\sqrt{2\pi}}\frac{s/k_B}{\sqrt{1-{\rm e}^{-2\gamma\delta\tau}}}
\exp\left[-\frac{s}{2k_B}\frac{\left(y_{2}-{\rm e}^{-\gamma\delta\tau}y_{1}\right)^2}{1-{\rm e}^{-2\gamma\delta\tau}}\right].
\label{dospuntos}
\end{equation}
Here we have defined the thermodynamical frequency $\gamma$,
\begin{equation} 
\gamma:=\frac{s}{R},
\label{klar} 
\end{equation}
with $R$ given as in (\ref{flujos}) and $s=-{\rm d}^2S/{\rm d}y^2|_0$. Furthermore, one can reexpress the probability density (\ref{dospuntos}) in terms of path integrals over thermodynamical configuration space: up to normalisation factors one finds \cite{XONSAGER}
\begin{equation}
f_1\left({y_{2}\atop \tau_{2}}{\Big\vert}{y_1\atop \tau_1}\right)
=\int_{y(\tau_1)=y_1}^{y(\tau_2)=y_2}{\rm D}y(\tau)\,
\exp\left\{-\frac{1}{2k_B}\int_{\tau_1}^{\tau_{2}}{\rm d}\tau\,{\cal L}\left[\dot y(\tau), y(\tau)\right]\right\}.
\label{camino}
\end{equation}
Above we have defined the thermodynamical Lagrangian function ${\cal L}$
\begin{equation}
{\cal L}\left[\dot y(\tau), y(\tau)\right]:=\frac{R}{2}\left[\dot y^2(\tau)+\gamma^2y^2(\tau)\right],
\label{lagratermo}
\end{equation}
whose actual dimensions are entropy per unit time.

\section{Irreversible thermodynamics {\it vs}\/. quantum theory}\label{sctres}

We can now establish a precise map between quantum mechanics and classical, irreversible thermodynamics. Let $t$ denote mechanical time, $m$ the mass of the quantum particle under consideration, and $\omega$ the frequency of a harmonic potential experienced by the particle.

In the first place, the thermodynamical time variable $\tau$ must be analytically continued into ${\rm i}t$:
\begin{equation}
\tau\leftrightarrow{\rm i}t.
\label{tiempo}
\end{equation}
Second, the thermodynamical frequency $\gamma$ becomes the mechanical frequency $\omega$ of the harmonic oscillator:
\begin{equation}
\gamma\leftrightarrow\omega.
\label{frecuencia}
\end{equation}
Next we map the thermodynamical variable $y$ onto the mechanical variable $x$:
\begin{equation}
y\leftrightarrow x.
\label{equisy}
\end{equation}
As a rule, $x$ will be a position coordinate. Hence there might be some dimensional conversion factor between $x$ and $y$ above, that we will ignore for simplicity. Bearing this in mind, we will finally make the identification
\begin{equation}
\frac{s}{2k_B}\leftrightarrow\frac{m\omega}{\hbar}
\label{identifizz}
\end{equation}
between thermodynamical and mechanical quantities. We have expressed all the above replacements with a double arrow $\leftrightarrow$ in order to indicate the bijective property of our map between quantum mechanics and classical, irreversible thermodynamics.

On general grounds, applying the replacements (\ref{tiempo}), (\ref{frecuencia}), (\ref{equisy}) and (\ref{identifizz}), one expects thermodynamical conditional probabilities to map onto mechanical conditional probabilities\footnote{While $f_1$ is a probability density, $K$ is a probability density amplitude; see ref. \cite{XNOI2} for a discussion of this issue.},
\begin{equation}
f_1\left({y_2\atop \tau_2}{\Big\vert}{y_1\atop \tau_1}\right)\leftrightarrow K(x_2,t_2\vert x_1,t_1),
\label{condprob}
\end{equation}
while thermodynamical unconditional probabilities are expected to map onto mechanical unconditional probabilities:
\begin{equation}
f_1\left({y\atop \tau}\right)\leftrightarrow \vert\psi(x,t)\vert^2.
\label{uncondprob}
\end{equation}
Here $K(x_2,t_2\vert x_1,t_1)$ denotes the quantum--mechanical propagator, and $\psi(x,t)$ is the wavefunction. As in (\ref{equisy}) above, one must allow for possible numerical factors between probabilities on the thermodynamical and on the mechanical sides; otherwise bijectivity is perfectly preserved.

Our expectations (\ref{condprob}), (\ref{uncondprob}) are borne out by experiment---experiment in our case being explicit computation. Indeed one finds the following. For $\gamma\to 0$, the irreversible thermodynamics corresponds to the free quantum--mechanical particle:
\begin{equation}
K^{\rm (free)}(x_2,t\vert x_1,0)=\sqrt{\frac{k_B}{s}}\,f_1\left({x_2\atop {\rm i}t}{\Big\vert} {x_1\atop 0}\right)_{\gamma\to 0},
\label{libre}
\end{equation}
while, for $\gamma\neq 0$, the irreversible thermodynamics corresponds to the quantum mechanics of a harmonic oscillator:
\begin{equation}
f_1\left({x_2\atop {\rm i}t}{\Big\vert}{x_1\atop 0}\right)=\exp\left(\frac{{\rm i}\omega t}{2}-\frac{\Delta V}{\hbar\omega}\right)\sqrt{\frac{2m\omega}{\hbar}}\,K^{\rm (harmonic)}\left(x_2,t\vert x_1,0\right).
\label{armonico}
\end{equation}
Above, $V=m\omega^2x^2/2$ is the harmonic potential, and $\Delta V=V(x_2)-V(x_1)$. Moreover, if $\psi_0(x)=\exp\left(-m\omega x^2/2\hbar\right)$ is the harmonic oscillator groundstate, then it holds that, up to normalisation,
\begin{equation}
f_1\left({x\atop {\rm i}t}\right)=\exp\left(-\frac{m\omega}{\hbar}x^2\right)=\vert\psi_{0}^{\rm (harmonic)}(x)\vert^2,
\label{onda}
\end{equation}
as expected.

{}Finally the path--integral representation of quantum--mechanical propagators,
\begin{equation}
K\left(x_2,t_2\vert x_1,t_1\right)=\int_{x(t_1)=x_1}^{x(t_2)=x_2}{\rm D}x(t)\,\exp\left\{\frac{{\rm i}}{\hbar}\int_{t_1}^{t_2}{\rm d}t\,L\left[x(t),\dot x(t)\right]\right\},
\label{prepa}
\end{equation}
has a nice reexpression in terms of classical, irreversible thermodynamics. Indeed,  applying our dictionary (\ref{tiempo}), (\ref{frecuencia}), (\ref{equisy}) and (\ref{identifizz}) to the mechanical path integral (\ref{prepa}), the latter becomes the thermodynamical path integral already seen in (\ref{camino}). This leads us to the following relation between the action integral $I$ of the mechanical system and the entropy $S$ of its thermodynamical counterpart:
\begin{equation}
\frac{{\rm i}}{\hbar}I\leftrightarrow\frac{1}{k_B}S.
\label{relazione}
\end{equation}
It should be remarked that both $I$ and $S$ independently satisfy an extremum principle. In the Gaussian approximation considered here, the respective fluctuations (measured with respect to the corresponding mean values of $I$ and $S$ as given by their extremals)  are obtained upon taking the exponentials. We thus obtain the quantum--mechanical
wavefunction and the Boltzmann distribution function:
\begin{equation}
\psi=\sqrt{\rho}\,\exp\left(\frac{{\rm i}}{\hbar}I\right), \qquad \rho_B=Z^{-1}\exp\left(\frac{1}{k_B}S\right).
\label{obtenemos}
\end{equation}
As usual, $Z$ denotes some normalisation factor. Since, by the Born rule, we must have $\rho_B=\vert\psi\vert^2$, this provides us with an elegant expression combining thermodynamics and quantum mechanics into a single equation:
\begin{equation}
\psi=Z^{-1/2}\exp\left(\frac{1}{2k_B}S\right)\exp\left(\frac{{\rm i}}{\hbar}I\right).
\label{elegantge}
\end{equation}

Eqs. (\ref{relazione}) and (\ref{elegantge}) are very inspiring, as they reveal a fundamental complementarity between the mechanical action integral (on the mechanical side) and the entropy (on the thermodynamical side). We will later on return to the complementarity between these two descriptions, a feature already foreseen by Prigogine \cite{XPRIGOGINE}. For the moment let us simply remark the following consequence of this complementarity, namely, the symmetrical role played by Planck's constant $\hbar$ and Boltzmann's constant $k_B$. This latter property, and the ensuing entropy quantisation, have been discussed at length in refs. \cite{XNOI1, XNOI2}.

\section{Emergence from irreversibility}\label{sccuatro}

It has been claimed that {\it quantisation is dissipation}\/ \cite{XNOI3, BLASONE0, BLASONE00, BLASONE1, BLASONE2, BLASONE3, XELZE1, XTHOOFT1, XTHOOFT2, XSTABILE}---this claim is central to the emergence approach to quantum mechanics. In more precise terms, the previous statement implies that quantum behaviour can be expected from certain deterministic systems exhibiting information loss. One could compare this state of affairs to the relation between (equilibrium) thermodynamics and (classical) statistical mechanics. Namely, information loss in a microscopic theory (statistical mechanics) arises as the result of averaging out over many degrees of freedom; the emergent theory (thermodynamics) contains less information than its microscopic predecessor.

Thanks to the map established in section \ref{sctres}, the picture presented here features quantumness as an intrinsic property of dissipative systems. Conversely, by the same map, any quantum system features dissipation. In our picture, irreversibility and quantumness arise as the two sides of the same coin, thus becoming {\it complementary}\/ descriptions  of a given system ({\it complementarity}\/ being understood here in Bohr's sense of the word). As opposed to the emergence property discussed above,  the two theories (quantum mechanics and irreversible thermodynamics) contain exactly the same amount of information. It is interesting to observe that closely related views regarding the complementarity between mechanics and thermodynamics were defended long ago by Prigogine \cite{XPRIGOGINE}.

Now it has been (rightly) pointed out that correspondence and emergence are not quite the same concept \cite{XHU1}. This notwithstanding, we can still argue that quantum mechanics continues to arise as an {\it emergent phenomenon}\/ in our picture. This is so because Boltzmann's dictum applies: {\it If something heats up, it has microstructure}\/. In other words,  every thermodynamics is the coarse graining of some underlying statistical mechanics. Thus the mere possibility of recasting a given theory in thermodynamical language proves that the given theory is the coarse--grained version of some finer, microscopic theory.

\section{Gaussianity}\label{sccinco}

As a technical remark, we should point out that we have worked throughout in the Gaussian approximation. On the thermodynamical side of our map this corresponds to the linear response theory; on the mechanical side this refers to the harmonic approximation. Within the regime of applicability of this assumption we can safely claim to have provided a rigorous proof of the statement that {\it quantum mechanics is an emergent phenomenon, at least in the Gaussian approximation}\/.

Using the fact that any potential can be transformed into the free potential or into the harmonic potential by means of a suitable coordinate transformation (as in Hamilton--Jacobi theory \cite{XMATONE1,  XMATONE2}), one would naively state that the Gaussian approximation is good enough to ``prove" that quantum mechanics is an emergent phenomenon also beyond the Gaussian regime. However, this ``proof" overlooks the fact that quantisation and coordinate changes do not generally commute. Therefore the previous reasoning invoking Hamilton--Jacobi can only be seen as a plausibility argument to support the statement that quantum mechanics must remain an emergent phenomenon
{\it also beyond}\/ the Gaussian approximation. There is, however, abundant literature  dealing with the emergent nature of quantum mechanics, {\it regardless of}\/ the Gaussian approximation, using techniques that are very different from those presented here, and with a spectrum of applicability that ranges from the smallest \cite{XELZE2, XHU2} to the largest \cite{KIEFER, PENROSE}. We will therefore content ourselves with the proof of emergence presented here, the expectation being that some suitable extension thereof (possibly using perturbative techniques) will also apply beyond the Gaussian approximation.

\vskip1cm
\noindent
{\bf Acknowledgements} J.M.I. would like to thank the organisers of the {\it Sixth International Workshop DICE 2012: Spacetime--Matter--Quantum Mechanics:
from the Planck scale to emergent phenomena}\/ (Castiglioncello, Italy, September 2012) for the invitation to present this talk, for stimulating a congenial atmosphere of scientific exchange, and for the many interesting discussions that followed.\\
\noindent
{\it O muse, o alto ingegno, or m'aiutate;\\
o mente che scrivesti ci\`o ch'io vidi,\\
qui si parr\`a la tua nobilitate.}\/---Dante Alighieri.

\end{document}